\documentclass{elsart1p}
\usepackage{graphicx}
\usepackage{amssymb}
\begin{document}
\begin{frontmatter}
%
%
%
\title{The weak structure of the nucleon \\from muon capture on $^3$He}
%
%
\author{Doron Gazit}
%
\address{Institute for Nuclear Theory, University of Washington, Box 351550, 98195, Seattle, WA, USA.}
\begin{abstract}
I use a microscopic calculation of the weak capture process $^3\rm{He}(\mu^{-},\nu_\mu)^3\rm{H}$, to extract and constrain the weak form factors of the nucleon, particularly the induced pseudo-scalar, and the second-class currents. The induced pseudo-scalar form-factor is found to agree with the prediction of chiral perturbation theory. The constraint found on the conservation of vector current (CVC) hypothesis is the tightest to date.
\end{abstract}
\end{frontmatter}
The weak structure of the nucleon is a reflection of the influence of quantum chromodynamics (QCD) on renormalizing the couplings of weak probes to the nucleonic current. Thus, its measurement can be used to constrain and check the theory at low energies. The structure of the current, transferring momentum $q^\mu$, is dictated by Lorentz invariance, including a part with polar-vector symmetry \cite{1}:
\begin{equation} \label{eq:V_IA}
\small{\hat{J}^V_{\mu}={  F_V(q^2) \gamma_\mu 
+ \frac{i}{2M_N}F_M(q^2) \sigma_{\mu\nu}q^\nu + \frac{g_s}{m_\mu}q_\mu }},
\end{equation}
and with axial-vector symmetry:
\begin{equation}\label{eq:A_IA}
-\hat{J}^A_{\mu} =  { G_A(q^2) \gamma_\mu \gamma_5
+ \frac{G_P(q^2)}{m_\mu} \gamma_5 q_\mu + i \frac{g_t}{2M_N}\gamma_5\sigma _ {\mu\nu}q^{\nu} }.
\end{equation}
Here $m_\mu$ and $M_N$ are the masses of the muon and nucleon, respectively.
The fact that the nucleon has a complex internal structure implies a deviation from the $V-A$ structure, i.e. $F_V/G_A \ne 1$, and induces non-zero scalar $g_s$, pseudo-scalar $G_P$, weak-magnetic $F_M$, and axial-tensor $g_t$, form factors. $G_A$ is well known, mainly from neutron $\beta$-decay rate. The similarity between the structure of the weak vector current, and the electro-magnetic current has led to the CVC hypothesis, which suggests that these currents are connected via a rotation in isospin space. Implicitly, this entails that the weak vector form-factors are identical to the electro-magnetic form factors, and that $g_s=0$, as it precedes a term which breaks charge conservation. The latter can be connected to the fact that the $g_s$ and $g_t$ terms have abnormal behavior under G-parity, the combination of charge conjugation and rotation in isospin space, broken only by the small difference between the up and down quarks masses. Thus, $g_s$ and $g_t$ are highly suppressed \cite{18}. The induced pseudo-scalar form-factor $G_P$ is rather poorly known, to an accuracy of about 15\% \cite{15}, as it vanishes for low-momentum processes.

The weak process in which a muon is captured by a nucleus has been known as a useful laboratory for testing weak currents in the nucleus, due to the relatively high momentum transfer in this process, $\hbar |\vec{q}|/c \sim m_\mu$, which allows probing of all weak form-factors \cite{2}. In this contribution, I report a recent use of the weak capture process $^3\rm{He}(\mu^{-},\nu_\mu)^3\rm{H}$, to extract and constrain the weak form factors of the nucleon \cite{3}. The amazingly accurate measurement of this decay rate $\Gamma^{\rm{exp}}=1496(4) \,\rm{Hz}$ \cite{4}, makes this reaction an ideal test-site for the weak structure of the nucleon. This, however, necessitates a description of the problem from its nucleonic degrees of freedom, and its dynamics without any free parameters. The latter has limited the ability of previous attempts to use the reaction to constrain the nucleonic weak form-factors \cite{5}. 

Here, I use a different approach to the problem, resulting in a parameter free evaluation of the capture rate. First, the bound states of the trinuclei are calculated using realistic nuclear forces, specifically the nucleon-nucleon Argonne $v_{18}$ (AV18) potential \cite{6} combined with the Urbana-IX (UIX) three nucleon force \cite{7}. The calculation, accomplished using the EIHH method \cite{8}, yields the experimental binding energies (within $20 \,\rm{keV}$) and charge radii, and compares well with other {\it ab-initio} methods \cite{9}. Second, the dynamics of the capture is described using chiral perturbation theory ($\chi$PT) to fourth order. $\chi$PT is an effective theory of QCD at low-energies, that allows a systematic perturbative expansion of the QCD Lagrangian\cite{10}. By describing the dynamics of the problem within $\chi$PT, a straight connection to the underlying theory is achieved. Up to fourth order in $\chi$PT, the weak current includes a single nucleon current of the form of Eqs.~(\ref{eq:V_IA}-\ref{eq:A_IA}). To this order, CVC holds, and $g_t=0$. The pion form-factor was calculated within $\chi$PT \cite{11}, to be $G_P (q^2=-0.954 m_\mu^2) =7.99(20)$. In addition to the single nucleon current, at third order in chiral perturbation theory, meson exchange currents appear. Contrary to old models, the constraints exerted by the chiral symmetry dictate the structure of these currents. To the calculated order, they include only one unknown parameter, originating in a contact term of two nucleons, coupled to an external lepton, and calibrated using the triton $\beta$-decay rate. This parameter free, hybrid approach, named EFT*, has had impressive success in the literature \cite{12}.

In principle, the nuclear forces should be derived from the same chiral Lagrangian as the currents, and to the same order. However, this inconsistency is not expected to change the results significantly due to two main reasons. First, in a previous study of the reaction, Marcucci {\it{et al.}} \cite{5} have shown that the capture is essentially independent of the nuclear force, as long as it describes correctly the binding energies of the trinuclei. Second, in a recent calculation of triton decay, within chiral perturbation theory, the short-range correlations of weak character were shown to be disentangled from the short range correlation in the wave functions \cite{13}. Thus, only the long range part of the forces affect the reaction. This part is similar in the chiral forces and in the AV18+UIX model. This fact is apparent in the weak dependence the current calculation has on changing the $\chi$PT cutoff (less than $0.2\%$), indicating that the needed degrees of freedom are taken into account.

The prediction of the current approach for the capture rate is
$\Gamma= 1499 (10)_{NM} (6)_{\mathrm{RC}} \,\rm{Hz}$, consistent with the experimental measurement $\Gamma^{\rm exp}_{stat}=1496(4)\rm$$\,\rm{Hz}$.
The first theoretical error is due to the dependence in the nuclear model, including the $\chi$PT cutoff dependence, and uncertainties in triton half-life and in the low-energy constants $\chi$PT cutoff. The second error is due to theoretical uncertainty in the electroweak radiative corrections calculated for nuclei \cite{14}. The radiative corrections, which increase the decay-rate by 3\%, were not taken into account in previous studies.
The resulting constraint on the induced pseudoscalar form factor, $G_P(q^2=-0.954 m_\mu^2) = 8.13 \pm 0.6$, is in very good agreement with the $\chi$PT prediction. This should be compared to the current experimental determination, $G_P(q^2=-0.88 m_\mu^2) = 7.3 \pm 1.2 $ \cite{15}.

The axial G-parity breaking term is constrained by the calculation to be $\frac{g_t}{g_A}  = -0.1   \pm 0.68$, comparable to the current experimental limit $|g_t|<0.3$ at $90\%$ CL \cite{16}, though still far from its theoretical determination, $\frac{g_t}{g_A}=-0.0152(53)$ \cite{17}. 

The extracted value for the CVC breaking form factor according to this work is  $g_s = -0.005 \pm 0.040$. This represents a significant improvement to the current limit $g_s=0.01 \pm 0.27$ 
\cite{18}. 

Summarizing, I have shown that the weak process $^3\rm{He}(\mu^{-},\nu_\mu)^3\rm{H}$, can be used to get prominent bounds on the weak structure of the nucleon. In particular,
the induced pseudo-scalar form factor is constrained to $\pm 8\%$, and agrees with $\chi$PT prediction\cite{11}, and the CVC hypothesis is confirmed to a new limit $|g_s|<0.045$. 
The calculation shows the potential of nuclear {\it{ab initio}} calculations of weak processes as quantitative tests for the weak structure of the nucleon, as well as other properties of QCD at low energy. 

This work was supported, in part, by DOE grant number DE-FG0200ER41132.

\end{document}